%% file: main.tex
\documentclass[conference]{IEEEtran}
\usepackage{cite}
\usepackage{amsmath,amssymb,amsfonts}
\usepackage{algorithmic}
\usepackage{graphicx}
\usepackage{textcomp}
\usepackage{xcolor}
\usepackage{fancyhdr}
\usepackage{csquotes}

\usepackage[normalem]{ulem}
\usepackage{framed}
\usepackage{graphicx}
\usepackage{subcaption}
\usepackage{algorithm}
\usepackage{url}
\usepackage[acronym]{glossaries}
\glsdisablehyper
\usepackage[binary-units=true]{siunitx}
\usepackage{listings}
\usepackage{tabularx}
\newcolumntype{L}[1]{>{\raggedright\arraybackslash}m{#1}}
\newcolumntype{C}[1]{>{\centering\arraybackslash}m{#1}}
\newcolumntype{R}[1]{>{\raggedleft\arraybackslash}m{#1}}

\usepackage{amsthm}
\newtheorem*{claim}{Claim}

\def\BibTeX{{\rm B\kern-.05em{\sc i\kern-.025em b}\kern-.08em
    T\kern-.1667em\lower.7ex\hbox{E}\kern-.125emX}}

\fancypagestyle{firstpage}{
  \fancyhf{}

  \fancyhead[C]{\normalsize{Submitted to arXiv}}
  \fancyfoot[C]{\thepage}
}  

\graphicspath{ {images/} }

\title{SeMPE: \underline{Se}cure \underline{M}ulti \underline{P}ath \underline{E}xecution Architecture for Removing Conditional Branch Side Channels}

\author{\IEEEauthorblockN{Andrea Mondelli\IEEEauthorrefmark{1},
Paul Gazzillo\IEEEauthorrefmark{2} and  Yan Solihin\IEEEauthorrefmark{3}}
\IEEEauthorblockA{Department of Computer Science\\
University of Central Florida\\
Orlando, FL\\
\IEEEauthorrefmark{1}\texttt{mondelli@knights.ucf.edu},
\IEEEauthorrefmark{2}\texttt{paul.gazzillo@ucf.edu},
\IEEEauthorrefmark{3}\texttt{yan.solihin@ucf.edu}}}

\begin{document}
\maketitle
\thispagestyle{firstpage}
\pagestyle{plain}

\newacronym{sblock}{SecBlock}{SecureBlock}
\newacronym{sexec}{SecExec}{SecureExecution}
\newacronym{sjmp}{sJMP}{SecureJump}
\newacronym{eosj}{eosJMP}{End-of-SecureJump}
\newacronym{pc}{PC}{Program Counter}
\newacronym{npc}{nextPC}{Next Program Counter}
\newacronym{scf}{SCF}{Side-Channel Free}
\newacronym{fact}{FaCT}{Flexible Constant-Time Programming Language}
\newacronym{isa}{ISA}{Instruction Set Architecture}
\newacronym{jbt}{jbTable}{Jump-Back Table}
\newacronym{jbb}{jb}{Jump-Back}
\newacronym{lifo}{LIFO}{Last-In-First-Out}
\newacronym{rob}{ROB}{Reorder Buffer}
\newacronym{vsb}{ValidSB}{Valid Secure Block}
\newacronym{sempe}{SeMPE}{Secure Multi-Path Execution}
\newacronym{sprefix}{SecPrefix}{Secure Execution Prefix}
\newacronym{llvm}{LLVM}{Low Level Virtual Machine Compiler}
\newacronym{shadow}{ShadowMemory}{Shadow Memory Locations}
\newacronym{cpi}{CPI}{Cycles Per Instruction}
\newacronym{cte}{CTE}{Constant Time Expression}
\newacronym{icache}{IL1}{Instruction Cache}
\newacronym{dcache}{DL1}{Data Cache}
\newacronym{l2}{L2}{Second-Level Cache}
\newacronym{smt}{SMT}{Simultaneous Multi-threading}
\newacronym{lrs}{LRS}{Lazy Register Spill}
\newacronym{ilp}{ILP}{Instruction Level Parallelism}
\newacronym{pl}{PL}{Programming Language}
\newacronym{sdbcb}{SDBCB}{the secret-dependent behavior of conditional branches}
\newacronym{archrs}{ArchRS}{Architectural Register Snapshot}
\newacronym{rat}{RAT}{Register Alias Table}
\newacronym{gc}{GC}{Global Checkpoint}
\newacronym{nt-path}{NT-Path}{Not Taken Path}
\newacronym{t-path}{T-Path}{Taken Path}
\newacronym{spm}{SPM}{Scratchpad Memory}

\input{Text/abstract}
\input{Text/intro}
\input{Text/bkgnd}
\input{Text/threat}

\input{Text/design}
\input{Text/discuss}

\input{Text/method}

\input{Text/evalu}
\input{Text/concl}

\bibliographystyle{IEEEtranS}
\bibliography{main}

\end{document}

%% file: Text/abstract.tex
\begin{abstract}

One of the most prevalent source of side channel vulnerabilities is the secret-dependent behavior of conditional branches (SDBCB). The state-of-the-art solution relies on Constant-Time Expressions, which require high programming effort and incur high performance overheads. 
In this paper, we propose SeMPE, an approach that relies on architecture support to eliminate SDBCB without requiring much programming effort while incurring low performance overheads. The key idea is that when a secret-dependent branch is encountered, the \acrshort{sempe} microarchitecture fetches, executes, and commits both paths of the branch, preventing the adversary from inferring secret values from the branching behavior of the program. To enable that, \acrshort{sempe} relies on  an architecture that is capable of safely executing both branch paths sequentially. Through microbenchmarks and an evaluation of a real-world library, we show that \acrshort{sempe} incurs near ideal execution time overheads, which is the sum of the execution time of all branch paths of secret-dependent branches. \acrshort{sempe} outperforms code generated by \acrshort{fact}, a constant-time expression language, by up to a factor of $18\times$.
\end{abstract}

%% file: Text/intro.tex
\section{Introduction}

As more computation is performed in the cloud, secure and private computation becomes more and more critical. Sharing of hardware resources in the cloud is crucial to keeping their utilization rate high, but it opens the way for side channel vulnerabilities where an application may leak secret data through the usage patterns it exhibits on the shared hardware. Applications that share a hardware resource can then observe the resource usage pattern to infer secrets.  

An important and prevalent source of side channels is the {\em \acrfull{sdbcb}}.  Code such as \mbox{\textbf{if} \textit{(secret)} \textit{\{if-path\}} \textbf{else} \textit{\{else-path\}}} reveals information to the attacker through secret-dependent differences in the performance characteristics of the two paths resulting from the conditional.  For instance, the leak is a \emph{timing channel} when two paths differ in execution time, a \emph{cache access channel} if the paths differ in cache access counts or occurrences, a \emph{memory access pattern channel} if the memory accesses occur to different addresses in the two paths, and a \emph{branch predictor channel} when the branch predictor state captures the past outcomes of the branch. Rather than designing an architecture support to close each different side channel, in this paper we propose an architecture that removes a sources of these side channels.  

Figure~\ref{fig:modularexponential} is classic example of a side-channel attack, the modular exponentiation routine from RSA public-key cryptography.
The secrets are the bits of the key ($e$), tested on line~4 ($e_i = 1$).
The attacker can indirectly infer the value of $e_i$ by observing the time it takes to execute the operation. Closing secret-dependent conditional branches as a source of side channels is critical and challenging: the routines have to be carefully rewritten manually to eliminate secret-dependent conditionals~\cite{patch4NSS, addedmore, aestiming, bearssl, codingrules, whynot, pornin_why_nodate}. 

\begin{center}
\begin{figure}[htbp]
\hfill%
\begin{subfigure}{0.35\textwidth}
\centering
\begin{small}
\begin{algorithmic}[1]
\FOR{$i=n-1$ to $0$}
\STATE $r \leftarrow square(r)$
\STATE $r \leftarrow modulo(r,m)$
\IF{$e_i = 1$}
\STATE $r \leftarrow multiply(r,b)$
\STATE $r \leftarrow modulo(r,m)$
\ENDIF
\ENDFOR
\end{algorithmic}
\end{small}
\end{subfigure}
\caption{Modular exponentiation in RSA with $e_i$ as secret.}
\label{fig:modularexponential}
\end{figure}
\end{center}

The large human resources involved in manually rewriting code means only the most sensitive software are protected, leaving secret user data in general-purpose applications unprotected.
Currently, there are several approaches to eliminating \gls{sdbcb}. A popular software technique, used in many cryptographic libraries, is to use \emph{\gls{cte}}. \Gls{cte} eliminates conditional statements by manually converting the conditions into arithmetic expressions used in the branch paths.

\begin{figure}[htbp]
\begin{footnotesize}
\begin{subfigure}{0.3\columnwidth}
\centering
\begin{algorithmic}[1]
\STATE \textbf{@secret} $A,B,C$
\IF{$A \lor B$}
\STATE $j \leftarrow j + 1$
\ELSE
\IF{$C$}
\STATE $k \leftarrow k + 1$
\ELSE
\STATE $k \leftarrow k - 1$
\ENDIF
\ENDIF
\end{algorithmic}
\caption{}%
\label{conditionalstatements}%
\end{subfigure}%
\hfill
\begin{subfigure}{0.6\columnwidth}
\centering
\begin{algorithmic}[1]
\STATE \textbf{@secret} $A,bA,B,bB,C,bC$
\STATE $bA \leftarrow (bool)A$
\STATE $bB \leftarrow (bool)B$
\STATE $j \leftarrow (bA \times bB + bA $
\STATE \quad \quad$ \times (1-bB) + (1-bA) \times bB) $ 
\STATE \quad \quad$ \times (j+1) + (1-bA)\times(1-bB)\times j $
\STATE $bC \leftarrow (bool)C$
\STATE $k \leftarrow (1-bA) \times (1-bB)\times bC \times (k+1)$
\STATE $k \leftarrow k+(1-bA)\times (1-bB) $
\STATE \quad \quad $\times (1-bC)\times (k-1)$
\end{algorithmic}
\caption{}%
\label{constanttime}%
\end{subfigure}%
\end{footnotesize}

\caption{Examples: (a) code with conditional
statements, and (b) its constant-time version. A, B, and C are secrets.}
\label{constattimecomparison}
\end{figure}
Figure~\ref{conditionalstatements} shows an example of a nested \texttt{if-else} statement that operates on secret user data $A$, $B$, and $C$.  Figure~\ref{constanttime} is the same program after a \gls{cte} transformation. Each secret condition ($A$, $B$, and $C$) is converted into a binary value ($bA$, $bB$, and $bC$). Each statement is converted into an expression that includes the logical combination of the binaries that produces the statement. For example, in line~3, $j$ is assigned the value of $j+1$ when $bA$ and $bB$ are true, $bA$ is true and $bB$ is false, or when $bA$ is false and $bB$ is true. Otherwise, when both $bA$ and $bB$ are false, $j$ is assigned its old value, i.e., it is not mutated.

Other approaches have also been proposed. Memory Trace Obliviousness (MTO)~\cite{Liu13:MTO} and GhostRider~\cite{liu15:ghostrider} transform code in order to equalize memory accesses in both branch paths and obfuscates their addresses using ORAM~\cite{goldreich_towards_1987,Ostrovsky:1990:ECO:100216.100289,Goldreich96softwareprotection}. Instead of equalizing the execution of both branch paths, Raccoon~\cite{rane_raccoon:_2015}, a software approach built on top of transactional memory hardware, executes both branch paths. Raccoon transforms code so that both branch paths are executed, converts every load and store to transactions (using transactional memory support), and relies on a conditional move instructions (\texttt{CMOV}) to ensure that true-path values are written to memory. 

We introduce \emph{\gls{sempe}}, an approach that extends existing microarchitecture to eliminate \gls{sdbcb}.
Table~\ref{tb:comparo} compares the three prior approaches with \emph{\gls{sempe}} across four categories important for protecting private user data in the cloud:: (1) \emph{programming complexity} to encourage its use, (2) \emph{low performance overhead} for real-world scalability, (3) \emph{architectural simplicity} to ease adoption, and (4) \emph{backwards compatibility} for binary compatibility with non-\emph{\gls{sempe}} architectures.  \emph{\gls{sempe}} provides the best tradeoffs between security and performance, while remaining backwards compatible and simple to program.

Like Raccoon, \gls{sempe} works by executing both paths of branch instructions to eliminate the secret-dependent behavior, thereby preventing an adversary from inferring secret values.  Unlike Raccoon, however, our approach uses new hardware extensions that require minimal compiler support. \gls{sempe} repurposes and builds on \emph{dual-path execution}, originally proposed for improving the performance of hard-to-predict branches by speculatively executing both paths of a branch. Similarly, \gls{sempe} fetches and executes all paths of a secret conditional branch.  But, unlike prior dual-path execution architectures, \gls{sempe} ensures that the execution of both paths is indistinguishable from running either path alone, thereby preventing a side channel leak of secret values.  Achieving this security property requires major differences in the architectural design compared to traditional dual-path execution: an indistinguishable memory access pattern, an execution order independent of the branch condition, and the commit of all instructions of both paths.


\begin{table}[t]
\begin{footnotesize}

\begin{tabular}{|L{1.4cm}|C{1.4cm}|C{1.22cm}|C{1.2cm}|C{1.5cm}|} \hline
\textbf{Aspects} &\textbf{CTE}&\textbf{GhostRider}&\textbf{Raccoon} &\textbf{\acrshort{sempe}} \\ \hline \hline
Approach & elim. cond. branch & equalize path & execute both paths & {\bf execute both paths}\\ \hline
Technique & SW & HW/SW & SW & {\bf HW/SW} \\ \hline
Programming complexity & High & Low & Low & {\bf Low} \\ \hline
Reported Overheads & $187.3\times$  & $1,987\times$ & $452\times$ & {\bf $10.6\times$}\\ \hline
Simple architecture & Yes & No & Yes & {\bf Yes} \\ \hline
Backward compatible? & Yes & No & No & {\bf Yes} \\ \hline
\end{tabular}

\end{footnotesize}
\caption{Comparing approaches to eliminate \gls{sdbcb}: constant time expression (CTE), GhostRider~\cite{Liu13:MTO,liu15:ghostrider}, Raccoon~\cite{rane_raccoon:_2015}, and our \gls{sempe} Architecture.}
\label{tb:comparo}
\end{table}

\gls{sempe} introduces a new branching instruction, the Secure Jump (sJMP).
When executed, the sJMP instruction pushes the destination address into a hardware \gls{lifo} structure. When all the subsequent instructions have been committed, the pushed address is popped and used to set the \gls{npc}, automatically executing the other branch of a secret-dependent conditional.

Rather than introducing a new opcode to use sJMP, the programmer or compiler prefixes a normal branch instruction with a special byte at the beginning and end of the branch.
For optimization purposes, a programmer can omit the byte to use a non-secure branch for code not working with secret values.
This design simplifies conversion of both hand-written and automatically generated assembly code.
In contrast, writing a \gls{cte} algorithm has been cited as notoriously difficult~\cite{patch4NSS, addedmore, aestiming, bearssl, codingrules, whynot, pornin_why_nodate, aciicmez_predicting_2006, fardan_lucky_2013, almeida16, bernstein_cache-timing_2005, brumley_remote_2005} and carries hefty performance overheads. While domain specific languages have been proposed to reduce the programming effort~\cite{cauligi_fact:_2017,Cauligi:2019:FDT:3314221.3314605}, such solutions require rewriting software in a new language.

The byte chosen for sJMP is ignored on non-\gls{sempe} architectures, enabling backwards compatibility of \gls{sempe} assembly code, therefore
\gls{sempe} code can run on existing architectures without modification, albeit without the same security guarantees. This is in contrast with Raccoon which requires processors that have hardware transactional memory support.  For a feasible security solution to have widespread adoption in commercial systems, the performance overhead should be minimal.  Both MTO and Raccoon report overheads of $195\times$ and $22\times$, respectively, on average, and $1,987\times$ and $452\times$ in the worst case~\cite{rane_raccoon:_2015}.  While that direct comparisons between these overheads with each other and with \gls{sempe} is not feasible due to differences in benchmarks and machine assumptions, our evaluation of \gls{sempe} found an overhead of only $10.6\times$ even in the case of conditionals nested ten deep and on a real-world case of a side-channel vulnerability.

We evaluated the performance of our proposed architecture with both a set of microbenchmarks and a real-world software library for image conversion called \emph{libjpeg}~\cite{noauthor_libjpeg_nodate} that contains a side channel leak that reveals an image visual details during decompression. The use of microbenchmarks allows for targeted stress testing of \gls{sempe} performance by controlling the number and nesting depth of multiple secret-dependent branches. The \emph{libjpeg} evaluation demonstrates \gls{sempe}'s ability to remove a side channel using a variety of image types of sizes.  Our evaluation shows that the execution time with \gls{sempe} is near ideal: execution time increases linearly with the number of secret branch paths, independent of the size of the workload executed.  When compared against \gls{cte} code derived using the state of the art \gls{cte} language and compiler (\acrshort{fact}), \gls{sempe} outperforms \gls{cte} substantially, by a factor of \mbox{$1.6-18\times$}.

The rest of the paper is organized as following. Section~\ref{sec:background} covers background and related work.  Section~\ref{threatmodel}  discusses the threat model. Section~\ref{sec:design} discusses the proposed architecture design, and Section~\ref{sec:discussion} limitations and compiler support.  Section~\ref{sec:simulator} and Section~\ref{sec:evaluation} discuss the evaluation methodology and evaluation results. Finally, Section~\ref{sec:conclution} concludes the paper. 

%% file: Text/bkgnd.tex
\section{Background and Related Work}
\label{sec:background}

Several techniques have been proposed for eliminating the secret-dependent behavior of conditional branches, including constant time expressions, memory trace obliviousness, and hardware transactional memory. \gls{sempe}, however, draws inspiration from multi path execution in order to provide high performance while providing security guarantees.

\subsection{Techniques to Remove SDBCB}
\label{sec:constanttime_background}

\paragraph{Constant Time Expression} Manual programming effort for \acrfull{cte} is currently standard practice technique for eliminating \gls{sdbcb}. While popular, it has two substantial drawbacks that limit its use to simple code structures, such as in some crypto libraries. First, it involves a large manual effort, because it prohibits programmers from using conditional statements that use secret data.  Moreover, programmers need to inspect the resulting assembly after each compilation to ensure that the compiler has not added conditional branches.  Compilers are known to have inserted conditional branches even when the source code contains know conditional constructs~\cite{cauligi_fact:_2017,Cauligi:2019:FDT:3314221.3314605}.
Writing a constant-time algorithm is notoriously difficult~\cite{aciicmez_predicting_2006,fardan_lucky_2013,pornin_why_nodate,brumley_remote_2005}.
Part of the reason is that the complexity of \gls{cte} code increases super-linearly with the nesting depth of conditional branches.  In response, a domain specific language, \gls{fact}~\cite{cauligi_fact:_2017,Cauligi:2019:FDT:3314221.3314605}, has been proposed to simplify \gls{cte} programming.  While simpler to program, substantial restrictions exist at least in the latest version, e.g. no manual memory allocation, no function pointers, no function calls, no floating point, etc. can be used.  \Gls{fact} is a new programming language, making it difficult to use for existing production software. Finally, performance overheads incurred by \gls{cte} are very high.  In Figure~\ref{conditionalstatements}, the original code contains three additions, but the constant-time version in Figure~\ref{constanttime} contains 28 additions or multiplications, nearly an order of magnitude higher.
Since \gls{cte} is standard practice today, we compare \gls{sempe} against \gls{cte}.

\paragraph{Memory Trace Obliviousness} 
Memory Trace Obliviousness~\cite{Liu13:MTO} and the compiler and architecture for it (GhostRider~\cite{liu15:ghostrider}) transform code in order to balance memory accesses in both branch paths and obfuscate their addresses using ORAM.  For example, if in the \emph{if} path there is an array access, then a new array access is added to the \emph{else} path. ORAM is used to randomize memory addresses so that the two array accesses are indistinguishable from the point of view of the address stream.

\paragraph{Raccoon}
Raccoon~\cite{rane_raccoon:_2015} is a software approach that uses hardware transactional memory to executes both branches of a secret-dependent conditional.  Raccoon works by modifying code so that both branch paths are executed, converting every load and store to a transaction, and relying on a \texttt{CMOV} instruction to ensure that the only true-path store value is written to memory.  While also executing both branch paths, \gls{sempe} does so directly through new hardware mechanisms without depending on code transformation to transactions and, moreover, does not incur the overhead of transactions wrapping every load and store.

\subsection{Multi Path Execution}

Dual/Multi Path Execution is a class of microarchitecture techniques previously proposed to reduce branch misprediction penalties by executing instructions from all paths of a conditional branch instruction~\cite{heil_selective_1996,Aragn2002DualPI,wallace_threaded_1998}. Once the branch outcome is discovered, the false path instructions are be squashed while the true path instructions are allowed to commit. The Dual Path Instruction Processing (DPIP)~\cite{Aragn2002DualPI} allows false-path instructions to be fetched, decoded, renamed, but not executed, while predicted-path instructions are executed. The Selective Dual Path Execution (SDPE)~\cite{heil_selective_1996} selectively forks a second path when a low confidence branch prediction is encountered. Threaded Multi-Path Execution (TME)~\cite{wallace_threaded_1998} allows the alternative path instructions to execute in a separate thread context of an \gls{smt} processor.

While the goal of previous multi-path execution architectures was to reduce the branch misprediction penalty for hard to predict branches, the goal of \gls{sempe} is to eliminate the \acrfull{sdbcb}.  Consequently, while sharing some similarities, \gls{sempe} is fundamentally different from traditional multipath execution in several ways. First, the execution of instructions from both branch paths must be indistiguishable to the observer in \gls{sempe}. That means that instructions from both paths must commit, instead of having one of them squashed.  Otherwise, the multi path execution may still leak a secret value.  Second, traditional multi path execution only handles one conditional branch, stalling at nested conditionals.  In contrast, \gls{sempe} must be able to handle nested conditional branches because a secure region may cover nested conditional branches that are both secret and non-secret.  Finally, the scope of traditional multipath execution is limited to the instruction window of the processor. In contrast, \gls{sempe} must handle an instruction count within secure conditional branch paths that often exceeds the processor instruction window. 

%% file: Text/threat.tex
\section{Threat Model}
\label{threatmodel}

We assume a threat model that is realistic for cloud computing where distinct applications share hardware. We assume that physical security is strong hence we do not protect against physical side channels (such as power usage) or other physical attacks. The victim and the attacker are assumed to run as separate processes in the same or different virtual machines that are scheduled to run on the same server, either in different cores sharing a cache, or in the same core through simultaneous multi-threading or time sharing.  

We assume that the hypervisor and OS are trusted, and that they correctly enforce address space isolation, so the attacker cannot directly read secret data of the victim.  We assume the attacker can measure timing at a coarse granularity, but has no access to hardware counters that track the victim's execution characteristics.  The attacker can prime the cache and branch predictor state through its own execution and can infer the victim's working set, i.e., addresses of past reads and writes to memory, through a shared cache.  The attacker knows or can guess the code that the victim is running.  We do not focus on eliminating specific side channels. Instead, we focus on eliminating a common source of various side channels: \gls{sdbcb}. We do not consider secret leaking through general memory access pattern.  If such leakage is present, we assume techniques such as Oblivious~RAM~\cite{goldreich_towards_1987} are used for protection, which are orthogonal to our work. We only seek to protect memory access patterns that leak a secret as a result of different conditional branch paths.  We note that \gls{sempe} does not address Spectre/Meltdown-style attacks~\cite{kocher_spectre_2018,lipp_meltdown:_2018},  because they do not involve leaks due to secret-dependent branch behavior.  Techniques for preventing Spectre/Meltdown are orthogonal to \gls{sempe}.

We rely on the same input program assumptions used by Raccoon~\cite{rane_raccoon:_2015}, i.e. (1) the program does not contain bugs that will induce application crashes, (2) the program does not exhibit undefined behavior, and (3) if multi-threaded, the program is data-race free.
Because the proposed architecture executes all paths of a secure branch, an instruction in a false path may incur an exception, such as due to operating on incorrect value (e.g. divide-by-zero). Such situations are normally acceptable even in a bug-free program, if the programmer assumed always-taken or always-not taken branch behavior for a specific secret.

%% file: Text/design.tex
\section{SeMPE Design}
\label{sec:design}

\subsection{Foundation of Security}

The foundation for security of SeMPE is that executing both paths of a conditional branch that depends on secret is necessary to hide the secret. Assume a conditional branch with the following form, \mbox{{\sf if (secret) P1 else P2 }}. Suppose that P1 and P2 {\em exclusive}, i.e. do not share common instructions, and {\em minimal}, i.e. removing any instruction from P1 (or P2) changes the live out values of P1 (P2). Also suppose that P1 and P2 are bug-free and do not incur any terminating exceptions. We claim that: 

\begin{claim}
For the secret to be not inferrable from the execution of P1 or P2, the minimum execution needed is all instructions of P1 and all instructions of P2.  
\end{claim}

To support the claim, consider the cases below. If only one of P1 or P2 is executed, secret is inferrable due to the behavior reflecting only one of them. If both P1 and P2 are executed entirely, secret cannot be inferred as execution behavior no longer depends on secret. Now suppose that we execute both P1 and P2 minus one instruction from P1. Since P1 is minimal, the correctness of P1 is affected. If P1 is the correct path, the execution of the code following the paths is affected and the change is observable by the attacker. If P1 is the wrong path, the execution of the code following the paths is not affected, but the observer expects change in behavior.  Hence, no instructions can be removed from P1 and P2. 

The important implication of the claim is that the execution time for execution of both paths of a secret branch represents the {\em ideal} overheads. If there are $N$-deep nested conditionals, and each path incurs $T$ time, the ideal execution time in theory is $2^N\times T$. Any secure execution must be evaluated against that ideal. 

\subsection{Terminology}

In order to be practical, \gls{sempe} design must meet the following criteria. First, the architecture modification to the processor core must be {\em simple} (low complexity). Second, it must be {\em bidirectionally backward compatible}: traditional code must run correctly on the new architecture, and modified code must run correctly on traditional architecture albeit without security guarantees. Third, it must incur {\em low programming effort} and preferably code transformation should be automatable. Finally, it needs to be {\bf fast}; {\em excessive} overheads are unlikely tolerable in production systems. To clarify the last point, the execution time must be as close as possible to the ideal case of the sum of execution time of all paths. 

\begin{figure}[t]
\centering
\includegraphics[scale=0.4]{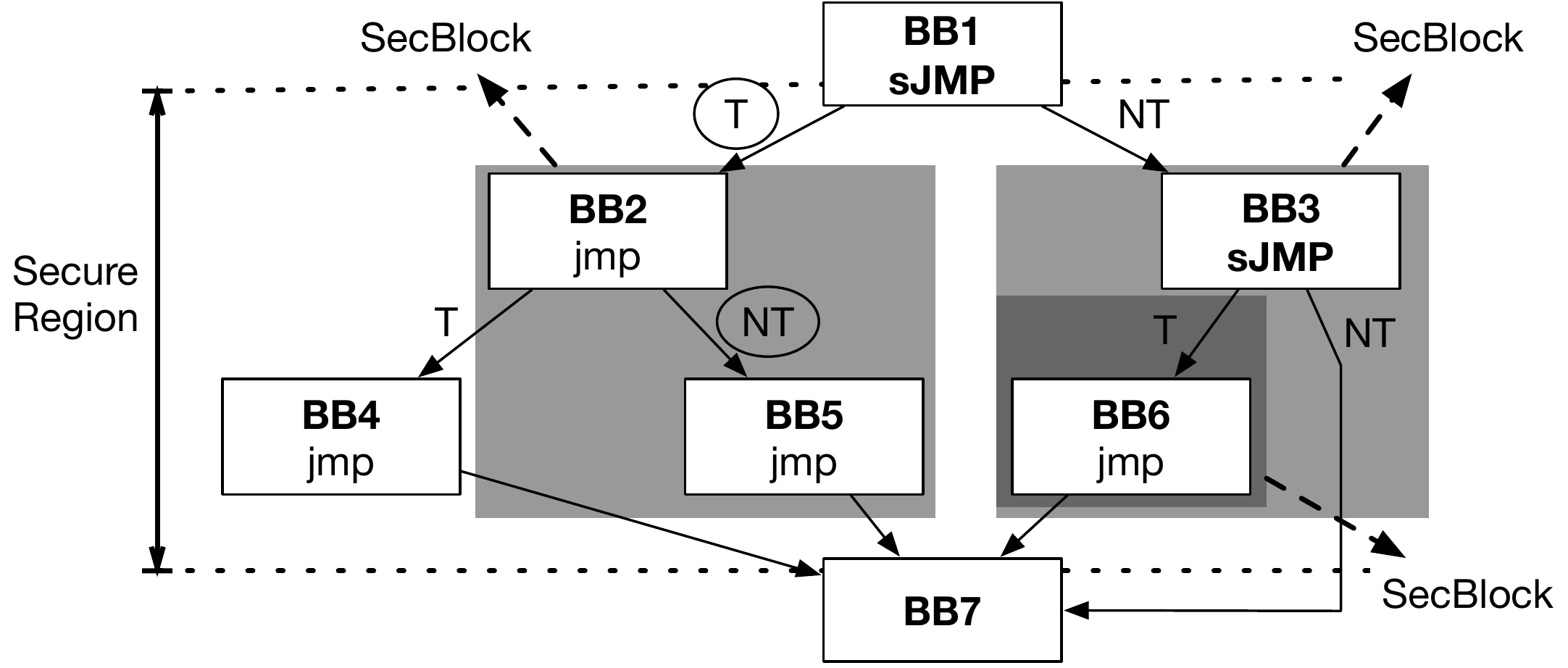}
  \caption{Illustrating key concepts used in \acrshort{sempe}: secure region, secure block, and secure branches.}
  \label{fig:bb}
\end{figure}

Before continuing, let us first discuss several terms. Suppose we have a control flow graph shown in Figure~\ref{fig:bb}, containing seven basic blocks. The true branch outcomes and paths are shown circled. Two basic blocks BB1 and BB3 contain secret-dependent conditional branches, denoted as sJMP. For convenience, we will refer them as \enquote{{\em secret branches}}. Other basic blocks contain either non-secret conditional branches or non-conditional branches, together denoted as {\sf jmp}. All instructions in the path of a secret branch are referred to as \gls{sblock}. The figure shows both paths of BB1's sJMP as \glspl{sblock}. In contrast, BB2's branch is not a secure branch hence BB4 and BB5 do not form \glspl{sblock}. \gls{sblock} can be nested, for example, BB6 is a \gls{sblock} contained within the larger (BB3, BB6) \gls{sblock}. The significance of \gls{sblock} is that all instructions in \gls{sblock} must be executed. For example, in the figure, the execution must cover BB1, BB2, BB5, BB3, BB6, and finally BB7. The only basic block that does not need to be executed is BB4, because it is not \gls{sblock}. The encapsulating (i.e. outermost) code starting from the secure branch to the joint point of its paths is referred to as the {\em secure region}. For a secret branch with two \glspl{sblock}, we refer to the true path as valid block.

\subsection{Expressing Secure Regions}

\paragraph{Instruction Set support}
\gls{sempe} needs to be able to identify a secure branch. To ensure backward compatibility, we add prefix to existing branch instructions instead of introducing entirely new branch instructions. For this discussion, we will assume  \mbox{x86\_64} \gls{isa}~\cite{corporation_intel_2016}, but a similar approach can be applied to other \gls{isa}s. The \mbox{x86\_64} is chosen because it was the most challenging to add new extensions and instructions, due to the variety and the large number of instructions~\cite{noauthor_x86_2019}. 

\lstloadlanguages{[x86masm]Assembler}
\lstdefinestyle{myAsm}{
language=[x86masm]Assembler,
    basicstyle=\fontsize{8}{6}\selectfont\ttfamily,
    tabsize=2,
    morekeywords={pn},
    deletekeywords={eax,mov,edx,jle},
    otherkeywords={2e,90,260,74,19,0x64,238}
}

To support the \gls{sempe}, the \gls{isa} is extended by adding a new instruction (eosJMP), and a unique prefix for branch instructions, called \gls{sprefix}. Branch instructions are coded as sJMP using the \gls{sprefix}. We use byte \textbf{0x2e}, which is normally interpreted as hints of static branch prediction to the compiler. 

The second modification is the addition of a new instruction that will be inserted as the first instruction in common between the two branch paths of the secure jump. The compiler inserts this instruction displacing the instruction that used to be the joint point of both branch paths. We refer to the new instruction as End-of-SecureJump (eosJMP). The instruction works as a backward jump to return the execution to the branch and the other branch path. We implement it using bytes \textbf{0x2e,0x90}. This instruction will be interpreted as a \texttt{NOP} in regular processors. 

By using prefix and \texttt{NOP}, the binaries that are backward compatible with regular processors, keeping the execution overhead-free when running on a legacy architecture. The instructions added are interpreted as secure branches only by the microprocessor described in this paper.


\subsection{Challenges to Multi-Path Execution}

\begin{figure}[t]
\centering
  \includegraphics[scale=0.4]{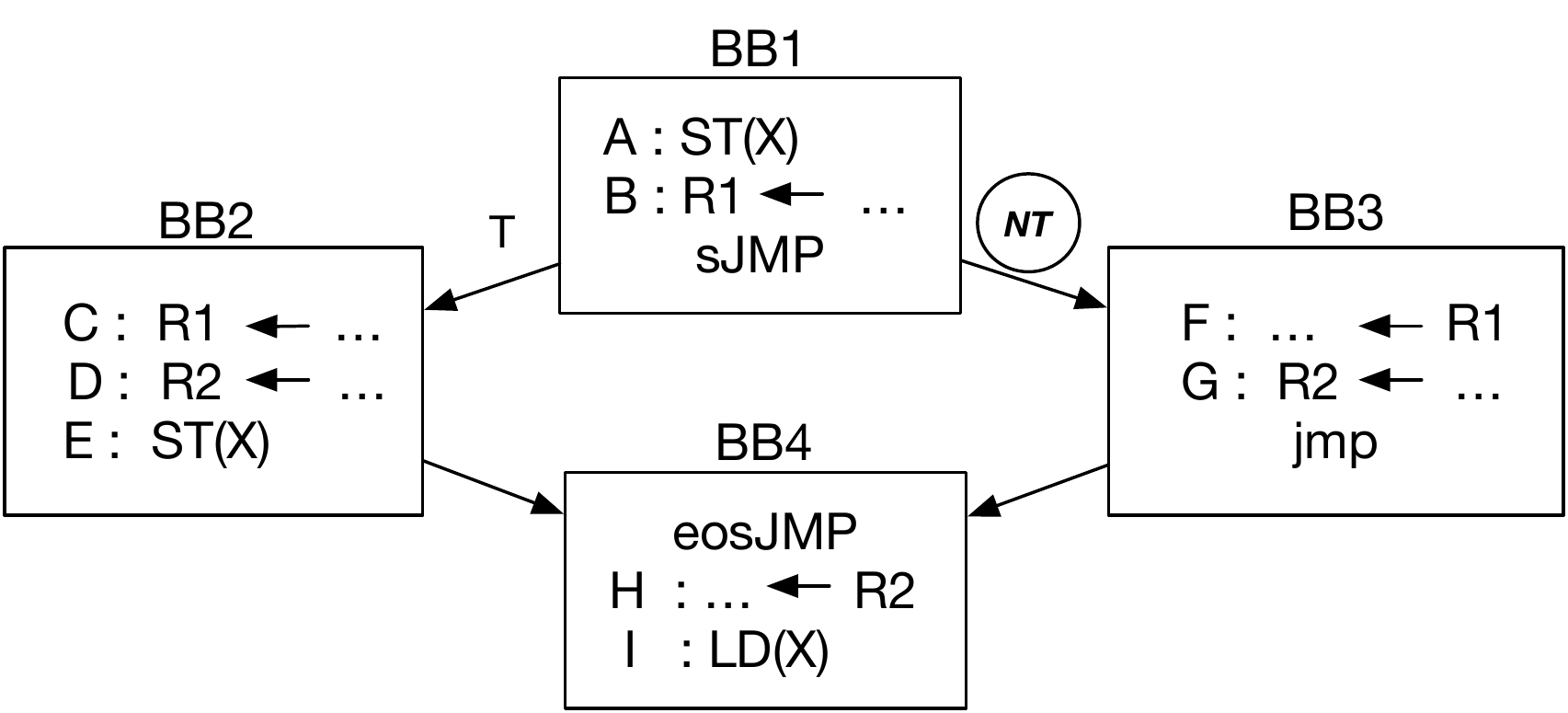}
  \caption{Basic Blocks with Phantom Dependencies. Secure branch's true path is not taken (NT). }
  \label{fig:falsedep}
\end{figure}

Multi-path execution introduces challenges in designing the pipeline. Consider a code example in Figure~\ref{fig:falsedep} with four basic blocks with several instructions in each basic block. Suppose that a secure branch's true path is not taken. Note that we have read after write (RAW) dependence between instructions B and F ($B \rightarrow_{RAW} F$), and between G and H ($G \rightarrow_{RAW} H$). If BB2 is also executed, phantom dependences may be introduced. An execution sequence of BB1, BB2, BB3, and BB4 will introduce the following phantom dependences: $B \rightarrow_{WAW} C$, $C \rightarrow_{RAW} F$, and $D \rightarrow_{WAW} G$. Likewise, if the execution sequence is BB1, BB3, BB2, and BB4, phantom dependences are also introduced. The dependences obviously affect the correctness of the execution. Phantom memory dependences are also possible, with $A \rightarrow_{MEM} I$ or $E \rightarrow_{MEM} I$ being phantom.  

Unlike past multi-path architectures, \gls{sempe}'s goal is not lowering branch misprediction penalty. A secret branch must execute and {\em commit} both branch paths regardless of the branch predictor. While instructions from both paths can be executed in parallel, it increase architecture complexity significantly, in particular the outcome of register renaming may be unpredictable due to phantom register dependences, and restoring state becomes complicated.   
To keep the hardware support simple, we choose to execute the paths sequentially: a secret branch is evaluated twice, as true for the first \gls{sblock} and as false for the second \gls{sblock}. 

Phantom dependences are still introduced with sequential execution of \glspl{sblock}. When a false-path \gls{sblock} is executed, the architecture state such as the rename table and register file will be changed. Thus, when eosJMP is encountered and the execution needs to go to the alternate path, the architecture state prior to the \gls{sblock} needs to be restored. Similarly, the architecture state corresponding to the true \gls{sblock} must be in place (or restored) prior to exiting the secure region. Section~\ref{sec:register} discusses our approach to this problem. 

A similar phenomena exists for memory dependences, except that memory values are not part of the micro-architectural state, so saving and restoring memory values is out of the scope of \gls{sempe}'s capabilities.  We assume that programs are written or compiled with memory dependences already disambiguated.

\subsection{SeMPE Microarchitecture}
\label{sec:microarch}


In this section, we describe the architecture to enable secure execution of both branch paths of a secret branch.
In traditional architectures, when a conditional branch instruction is encountered, the \gls{npc} is set to either the following the instruction (if the branch is not taken) or the target branch address (if the branch is taken). The branch predictor outcome sets the \gls{npc} based on the predicted outcome. 

In \gls{sempe}, sJMP must execute both paths, hence the branch predictor does not need to generate prediction. Hence, the \gls{npc} is set to the following instruction address, as if the branch condition is not verified. 
The not-taken \gls{sblock} is executed entirely, while the target address of the sJMP instruction is calculated. Once the target address is calculated, it will be saved and used by the eosJMP instruction to set up the \gls{npc}, which corresponds to the first instruction in the second \gls{sblock}. \underline{Not-taken path is always executed first} hence no secret-dependent behavior can be observed by the attacker, including order of memory accesses and behavior of prefetcher. We also assume the attacker does not alter the code.

\begin{figure}[t]
\centering
\includegraphics[width=1\linewidth]{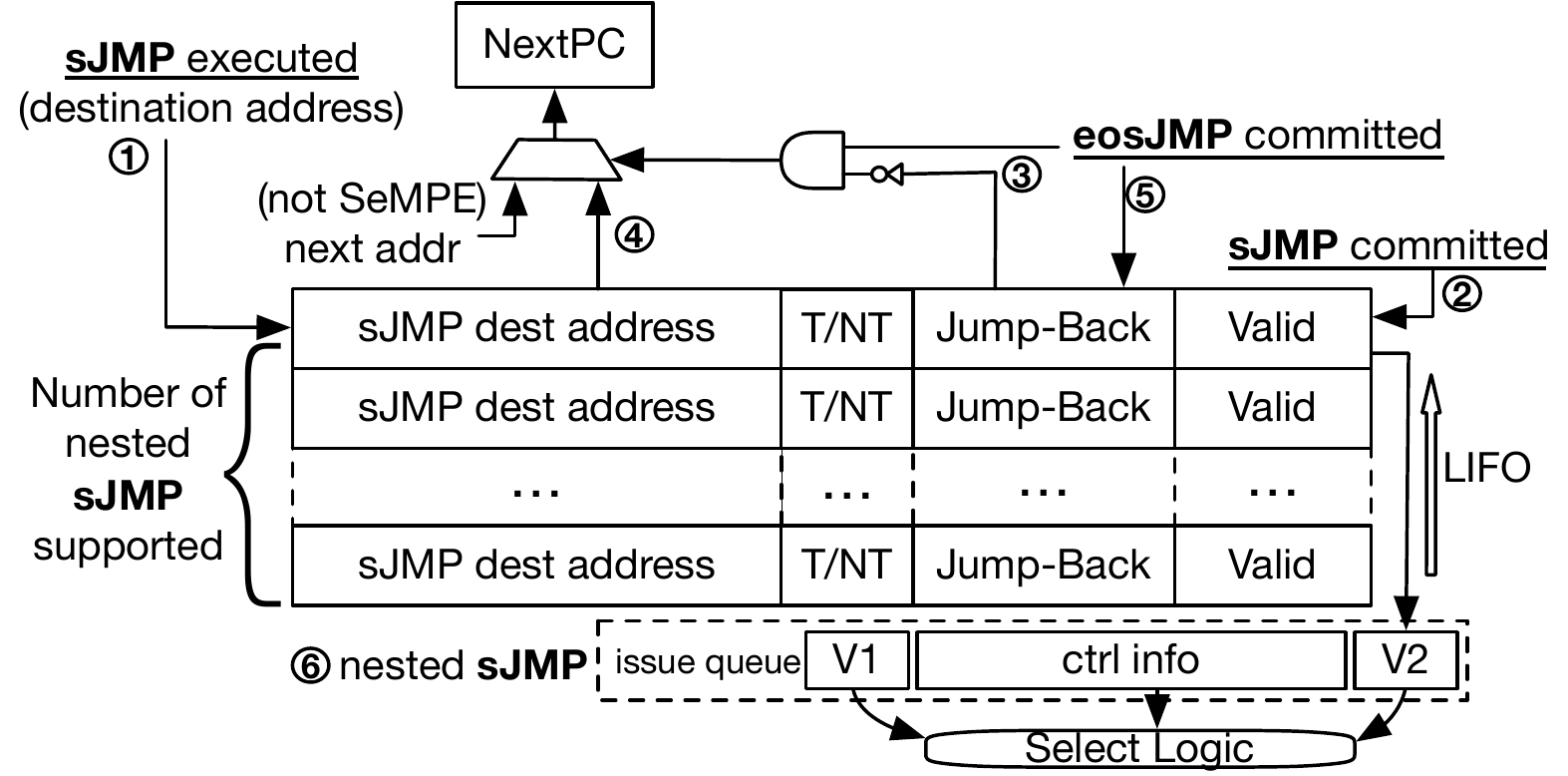}
  \caption{Micro-architecture support for \gls{sempe}. The branch outcome is saved in the T/NT bit field, where T  is $Taken$ and NT is $ Not Taken$}
  \label{fig:lifo}
\end{figure}

The target address is managed in a \gls{lifo} hardware structure, called a \gls{jbt}, shown in Figure~\ref{fig:lifo}. The \gls{jbt} consists of multiple entries to support nested secret branches, with each entry containing the \gls{npc} address, the branch outcome (T/NT), a valid bit (Valid), and a \gls{jbb} bit. When a sJMP is issued (Step \textcircled{1}), a new entry in the \gls{jbt} is created, with the Valid and \gls{jbb} reset. 
When the sJMP is committed, the calculated target address is written to the \gls{jbt}, and the Valid bit is set (Step \textcircled{2}). A sJMP instruction can only be issued if the prior \gls{jbt} entry has its Valid bit set, otherwise it must stall from issuing. In this way, the \gls{jbt} will be faithful to \gls{lifo} to ensure that the the correct Valid bit is set for the correct sJMP.

At the end of the first \gls{sblock}, the eosJMP is executed and committed (Step \textcircled{3}). At that time, the most recent \gls{jbt} entry is looked up. If the \gls{jbb} is not set (when the eosJMP is encountered for the first time), the address field of the most recent entry is copied to the \gls{npc}  (Step \textcircled{4}), and the \gls{jbb}  is  set  (Step \textcircled{5}). If, instead, the \gls{jbb} is already set, this indicates that the second \gls{sblock} of the sJMP has been executed and the corresponding entry of the \gls{jbt} can be removed.

The existing issue queue presents a valid bit for each source operand, called V1 and V2~\cite{doi:10.2200/S00309ED1V01Y201011CAC012}. In the simplified issue queue entry in Figure~\ref{fig:lifo}, assuming two source operands, the V1 and V2 bits are set when the corresponding operand are ready, or ignored when the operand is not used. In existing microarchitectures, the V2 bit remains unset for conditional branches and not used. \gls{sempe} set it when the Valid bit of the \gls{jbt} is set. A nested sJMP can be issued (Step \textcircled{6}) only if the \gls{jbt} is empty or the last sJMP in the \gls{lifo} is executed, i.e., the Valid bit is set and copied in V2. We don't need to modify the existing issue queue.

The use of a \gls{lifo} structure allows the handling of nested sJMP with low hardware complexity, without the need of a more complex random-access structures, and without  adding address comparison logic. 
When running a \gls{sblock}, we may encounter non-secret branch instructions. In contrast to sJMP instructions, they will consult (and update) the branch predictor. 

The sJMP does not need to use the branch predictor, because we know in advance we will execute both path despite the value of the secret. If the pipeline is flushed due to a branch misprediction, the flushing works as follows. For each sJMP squashed in the \gls{rob}, from the newest to the oldest, the most recent \gls{jbt} entry is deleted. The \gls{rob} will  contain, at any time, the sJMP instructions representing \gls{sblock} whose \gls{pc} has not \enquote{jumped back} yet, i.e. \gls{jbb} is still invalid. Since the address contained in the \gls{jbt} will be used as \gls{npc} only when the eosJMP is committed the first time, we can guarantee the correctness after the pipeline flush.

Since each entry of the table deals with one sJMP instruction in a secure region, the number of \gls{jbt} entries is equal to the maximum number of nested sJMP the architecture can handle. 

The total size of \gls{jbt} is small. Each \gls{jbt} entry equals to the size of a register (64 bits) $+$ two bits (\gls{jbb} and Valid bits). Even with 30 entries, \gls{jbt} has less than 256 bytes. We believe a few dozen entries should be sufficient, because outside of recursion, deeply nested secure branches are rare. Our investigation reveals that the  degree of  sJMP nesting on a cryptographic algorithm is likely much less than a dozen.  Dealing with secret user data may require a higher nesting degree, but unlikely to be beyond 30 in most situations.

Furthermore, the compiler can reduce the nesting degree by collapsing multiple conditionals into a single one with larger expression. For example, \mbox{{\sf if (A) \{if (B) ...\}}} can be converted into \mbox{{\sf if (A and B) \{...\}}}. Recursion may be either rejected at compile time, or made to trigger exception at run time. It is up to the exception handler whether to stop program execution, or to continue execution of the branch as non-secure. We note that such restrictions are also common in CTE. 

\subsection{Dealing with Phantom Register Dependences}
\label{sec:register}

\begin{figure*}[t]
\centering
  \includegraphics[width=1\linewidth]{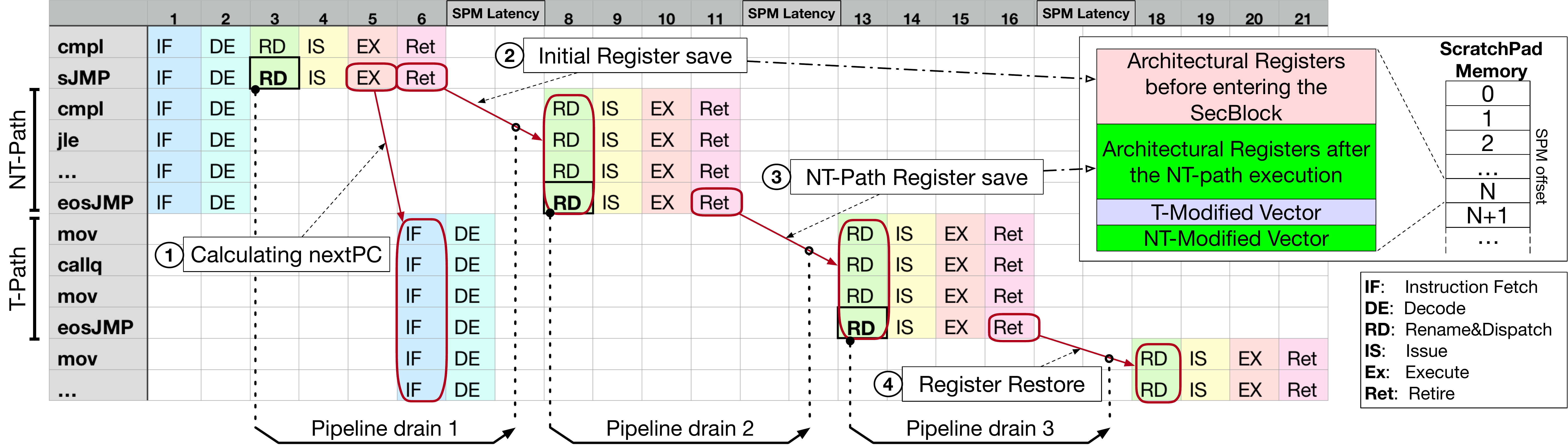}
  \caption{Example of SeMPE pipeline. Before entering the SecBlock and after the \mbox{NT-Path}, the pipeline is drained and the instruction rename is stopped. The pipeline is also drained at the end of the \mbox{T-Path}. The address of the first \mbox{T-Path} instruction is available after the sJMP execution~\textcircled{1}. The SPM snapshots contain the registers saved before entering the SecBlock~\textcircled{2} and after the \mbox{NT-Path}~\textcircled{3}. The two bit-vector, updated during the execution, are used to restore the correct register value at the end of both paths~\textcircled{4}. The address of the first instruction in the \mbox{T-Path} is available after the sJMP execution.}
  \label{fig:pipeline_ooo_with_spm}
\end{figure*}

Phantom register dependences are false register dependences that occur between both paths of a secure branch. To manage them, we consider several architecture solutions. The first solution considered was the Lazy Register Spill (LRS). 
LRS uses a cache-like rename table with tags, similar to~\cite{oehmke_how_2005}. The tag identifies the SecBlock, allowing to spill only modified registers. Unfortunately, LRS complicates the rename table and affects instructions not belonging to SecBlock. Our goal is to keep hardware changes low without impacting the performance of the rest of the program.
The second technique we considered was the use of a Physical Register Snapshot (PhyRS) mechanism to restore the contents of the register file and the \gls{rat} at the end of both paths, depending on the secret.

The implementation needs two snapshots per nested SecBlock, containing the register file and the Register Alias Table (RAT). The first snapshot is taken prior to the execution of a SecBlock, right after the sJMP is committed. The second snapshot is taken at the end of the execution of the not-taken path, when the eosJMP is committed for the first time. 
At the end of the SecBlock,  the register file and the RAT are rebuilt  using the correct snapshot, according to the branch outcome. For saving snapshots, we considered the combined of scratchpad memory and register spilling. The Scratchpad Memory (SPM) was used as a temporary buffer to mitigate register spilling before any nested SecBlock. 

This solution solves the problem of false register dependences between paths but introduced an excessive performance overhead during the memory spilling of the content of the SPM. In modern architecture, it is common to have hundreds of physical registers~\cite{amd_registers}. Saving all physical registers and the RAT~\cite{Xiao2013ImplementingFR} produce too much snapshot spilling to memory, especially for deeply nested conditional branches. 

Therefore, we choose a third design based on Architectural Register Snapshot (ArchRS) mechanism instead. The main difference is that only architectural registers are saved in the \gls{spm}, the number of which is much lower than the physical registers. Figure~\ref{fig:pipeline_ooo_with_spm} shows the composition of a  SecBlock snapshot at nesting level $N$. The nesting level is used as an offset to access the SPM during saving and restore. 

Along with the two architectural register states, one before entering the SecBlock and another after the \mbox{NT-Path} execution, the \gls{spm} contains two bit-vectors. Each vector contains many bits to the number of architectural registers. The vectors track the architectural register modified during the two paths, Taken Path (T-Path) and NotTaken Path (\mbox{NT-Path}), and will be used to restore the correct content of the architectural register at the end of SecBlock.
A  pipeline drain is added at the beginning of SecBlock. All the registers are saved when the sJMP is committed, and only modified registers are saved when the first eosJMP is committed. After the \mbox{NT-Path} the contents of the registers are restored from SPM. After the \mbox{T-Path}, the content of the architectural registers is updated with the correct value according to the secret. 


At the end of a SecBlock, the register restore phase takes place. The registers modified in at least one of the two paths are read from the SPM. Depending on the branch outcome contained in the corresponding jbTable entry, the register is overwritten with the correct value. 
Figure~\ref{fig:pipeline_ooo_with_spm} shows the sequence of executed instructions and when registers are saved or restored. The order of execution is independent of the secret. When the \mbox{NT-Path} is the true path, the value restored depends by the bit-vectors. For register modified in the \mbox{NT-Path}, the correct value comes from the \mbox{NT-Path} snapshot. 
For register modified in the \mbox{T-Path} but not in the \mbox{NT-Path}, the correct value is the one saved before entering the SecBlock. When the \mbox{T-Path} is the true path, all the modified register values are still read by the SPM but not used to restore the register contents. Instead, the current value is overwritten by itself. This behavior prevents the attacker from deducing the secret with a timing attack~\cite{percival_cache_2005,bernstein_cache-timing_2005,brumley_remote_2005,zhang15,Luo:2019:STA:3341169.3341729}.

The execution of secret blocks is never interleaved, so one secret branch is always completely executed until the eosJMP, which occurs just before the \texttt{CMOV}. The pipeline is drained after each eosJMP. This pipeline drain allows that (1)  the instruction window does not contain instructions from both paths at the same time, and (2) the instructions after the SecBlock observe the correct state of memory and registers. 

The ArchRS mechanism introduces a third pipeline drain before entering the SecBlock, so that only the contents of valid registers are saved without introducing an additional level of complexity in the reconstruction of the RAT. This pipeline drain is less expensive than a normal branch misprediction because the instructions are still fetched and decoded correctly, until their queues are full.
Registers modified in at least one of the two paths are always read by the \gls{spm}, even if  not used to restore the corresponding register value.

\subsection{Security Analysis}
\gls{sempe} eliminates \gls{sdbcb} through the execution of both branch paths (\glspl{sblock}) in an order not related to the secret.  The branch predictor channel, where the branch predictor state captures the past outcomes of the branch, is eliminated since there is no use of the predictor branch for sJMP. The compiler needs to reject any \glspl{sblock} that have a potential hardware exception, e.g., a divide-by-zero error, removing any potential leaks due to exceptions.  The user can decide whether to risk such code or not.

The combined use of \gls{shadow} and \texttt{CMOV} hides the cache access to the attacker. The attacker, therefore, is not able to leak secret through the cache utilization analysis.

%% file: Text/discuss.tex
\section{Discussion and Limitations}
\label{sec:discussion}

\subsection{Phantom Memory Dependences}
\label{sec:memory}

Executing both paths of a branch may cause the same memory locations to be written or read, creating phantom memory dependences, which are more difficult to address because memory values are beyond the architecture state of the pipeline. A store cannot be rolled back easily once it has been committed. 
To obtain the effects of fully executing and committing both paths of a secure branch, we considered several solutions. First, we could design the cache to keep versions of data from taken and not-taken paths and discard one of them at the conclusion of secure branch execution. However, keeping multiple versions of data in the cache creates complication with addressing and cache coherence, since one address may correspond to multiple data values. Furthermore, the manner in which values are discarded in the cache may cause a new side channel if not implemented carefully. 

To keep SeMPE simple, to deal with phantom memory dependences, we duplicate any memory-allocated data modified in the SecBlocks for each secret branch, disambiguating the memory and preventing conflicting reads and writes to the same memory location by the false path. We refer to the duplicated memory as \gls{shadow}.
At the joint point (i.e. the postdominator block), a conditional move instruction \texttt{CMOV} is used to copy one of the values to the original copy.
This completely avoids having to depend on a memory snapshot or use a memory state rollback. The approach has some similarity with Raccoon~\cite{rane_raccoon:_2015}, but with substantial differences: (1) we apply this only to memory locations, as phantom register dependences are handled differently (Section~\ref{sec:register}), and (2) we do not use the transactional memory and transaction buffers that Raccoon uses. As a result, SeMPE overheads come only from the expansion of memory footprint due to privatization, and the instruction execution overheads that come from executing \texttt{CMOV}s. 

\subsection{Compiler Support for \gls{sempe}}

The benefits of \gls{sempe} depend on correct usage of the \gls{isa}'s two new instructions, SecureJump and End-of-SecureJump.  These instructions mark the beginning and end of secure branches due to conditional branches on secret values.
Such usage can be automated in the compiler, however, using a combination of information flow algorithms that track secrets and existing control- and data-flow analyses available in modern compiler frameworks, e.g., LLVM.

Using \gls{sempe} correctly requires identifying the branches of secret values.  Automatic identification is possible by leveraging existing work on information flow analysis~\cite{ml-ifc-97,sabelfeld03,smith07,planul13,rodrigues16,king08}.  Information flow can be used to check for leaks of secret values from a \emph{source}, e.g., input from a protected database, to a non-secret \emph{sink}, e.g., an attacker-accessible output channel.  In the case of secret-dependent branches, the sinks are all branch statements.

Once the compiler has identified which conditional branches involve secrets, the compiler can identify which basic blocks of the control-flow graph are the secure blocks.  The secure blocks are successors of blocks that have secret-dependent branches, e.g., BB2, BB3, and BB6 from Figure~\ref{fig:bb}.  These secure blocks can then be transformed automatically for use with \gls{sempe}.  The compiler need only insert the secret-dependent branch with an sJMP where it would normally insert a JMP, and insert a eosJMP at the join point of the branch's two paths.
For instance, in Figure~\ref{fig:bb}, the end point for the sJMP in BB3 is the beginning of BB7, the first point after finishing the execution of any the resulting secret blocks.
In a control-flow graph, this point is the immediate postdominator of BB3, i.e., the first block through which any path from BB3 must enter~\cite{Muchnick:1998:ACD:286076}.

For a single secret-dependent branch, the eosJMP will end up being the successor of all secret blocks due to a secret-dependent branch.
Conditional statements may have nested conditionals, either secret or non-secret.  BB3 in Figure~\ref{fig:bb} is due to a secret branch nested inside of the branch resulting in BB1.  Each branch has its own set of secret blocks.  For each, the immediate postdominator indicates where the insert the eosJMP, handling the effects of any nesting.  In this case, both BB3 and BB1's postdominator is the same block, BB7.  While the postdominator need not be the same in all cases, one eosJMP is needed per sJMP.  In this case, when the postdominator is the same, the compiler needs to insert both \glspl{eosj} at the beginning of BB7. 

The \gls{isa} and its accompanying system software (assemblers, linkers, etc), require very little change.  Only two additional instruction types are needed.  The sJMP to indicate a jump into a secure block and eosJMP to indicate the end of a secure block.  The sJMP instructions have the same semantics as \texttt{JMP}, and are merely a signal to the hardware that both sides of the branch should be executed.  The eosJMP is equivalent to a \texttt{NOP} and is a signal that the secure block is complete.  The compiler toolchain need only emit these mnemonics and assemble them into the appropriate machine code.

A compiler can also help automate \gls{sempe}'s requirement of memory disambiguation in some cases.
The simplest solution for stack-allocated variables is to create additional stack frame entries for each variable used in both branches an sJMP.  A conditional move (\texttt{CMOV}) can then be inserted to select between the copies of the variable, as is done with hand-written \gls{sempe} code.

As with stack-allocated variables, phantom dependences can occur between two secret blocks due to accesses to the same heap location.
Such dependences are more difficult to detect at compile-time precisely and in general, in particular, for complex heap structures and pointer arithmetic.
There are some solutions to handling phantom heap dependences that could be employed to ensure correct usage of \gls{sempe}, e.g., shape analysis~\cite{Magill:2010:ANA:1706299.1706326,Hearn:2019:SL:3310134.3211968}, but may be too imprecise for some programs.  In the worst-case, a library for intercepting memory referencing could ensure disamguation at runtime, albeit at the expense of substantial overhead.

%% file: Text/method.tex
\section{Evaluation Methodology}
\label{sec:simulator}

To evaluate our scheme, we use two sets of workloads: microbenchmarks and a real-world application. The microbenchmarks are designed to stress test \gls{sempe} across a wide range of code characteristics. They are also useful due to the scarcity of real world applications that have been implemented with \gls{cte}; \gls{cte} is currently only used in crypto libraries. 

\begin{figure}[thbp]
  \includegraphics[width=1\linewidth]{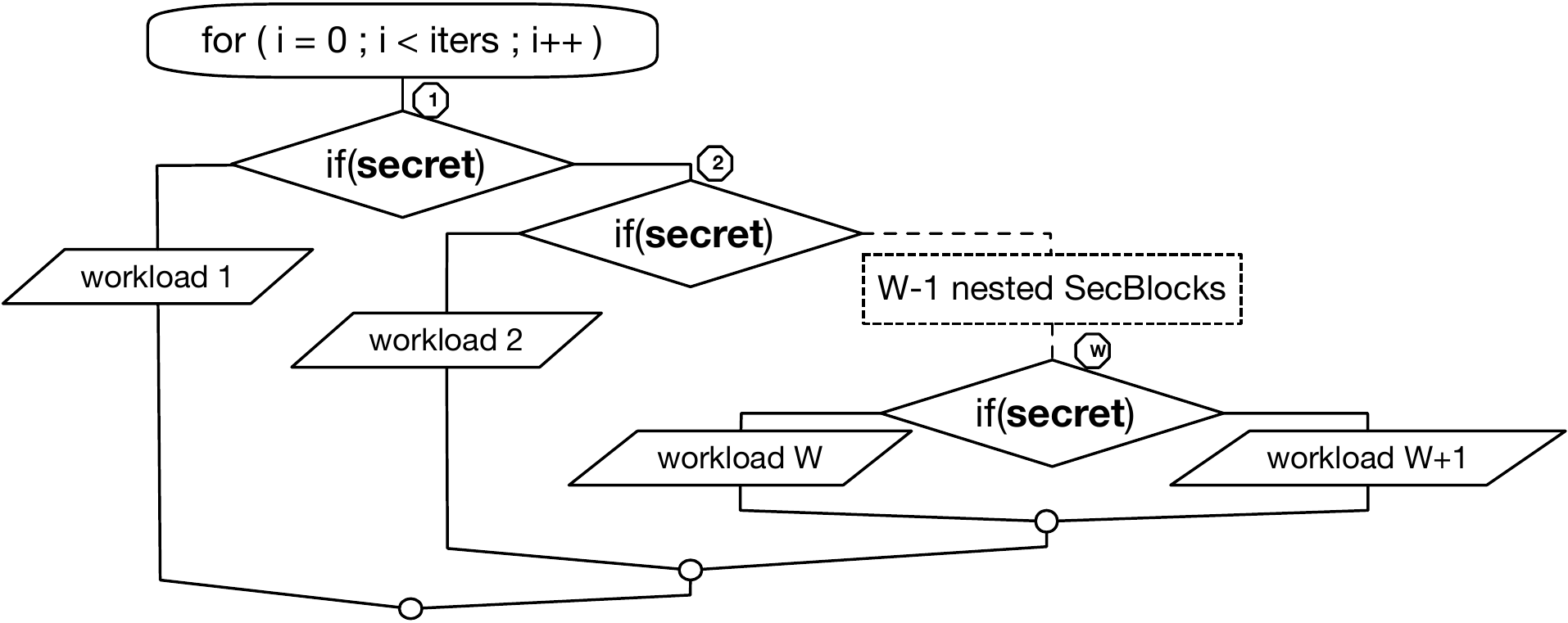}
  \centering
  \caption{Microbenchmark structure. The number of nested sJMP depends by $W$. The total number of sJMP per iteration is $W$, and the number of of nested sJMP is $W-1$ }
  \label{fig:microbech_structure}
\end{figure}

The microbenchmark has a customizable nested conditional branches that depend on secret, with several different workloads, as shown in Figure~\ref{fig:microbech_structure}. The workload is one of the following: 1) \textbf{Fibonacci}, which calculates Fibonacci series up to the specified term, 2)  \textbf{Ones}, which allocates a vector of integer, filling it with random numbers, and deleting the vector on exit, 3) \textbf{Quicksort}, which utilises a  \mbox{divide-and-conquer}  strategy to sort a large array~\cite{hoare_algorithm_1961},  and 4) Eight \textbf{Queens} problem~\cite{bell_survey_2009}, which places eight queens on an $8 \times 8$ chessboard such that none of them attacks one another. As can be seen in Figure~\ref{fig:microbech_structure}, the two main parameters of the microbenchmark are (1) the number of iterations of the entire secure region (\emph{I}) and (2) the nesting depth and width of each iteration (\emph{W}). We vary I and W to produce over 700 combinations in order to test the effect of nesting depth and eliminate measurement noise. We report a range of configurations for nesting depth: from $W=1$ (1-deep) to $W=10$ (10-deep). In all configurations, the number of instructions executed is at least $100$ million, run to completion. 

\begin{table}
\footnotesize 
  \centering
    \begin{tabular}{|l|l|}
    \hline
 clock frequency & 2.0 GHz \\ \hline
branch predictor & 31KB TAGE~\cite{seznec_new_2011}, 6KB ITTAGE~\cite{seznec_64-kbytes_2011} \\ \hline
fetch & 8 instructions / cycle \\ \hline
decode & 8 $\mu$ops / cycle \\ \hline
rename & 8 $\mu$ops / cycle \\ \hline
issue (micro-ops) & 8 $\mu$ops  \\ \hline
load issue & 2 loads / cycle \\ \hline
retire & 12 $\mu$ops / cycle \\ \hline
reorder buffer (ROB) & 192 $\mu$ops \\ \hline
physical registers & 256 INT, 256 FP \\ \hline
issue buffers & 60 INT / 60 FP $\mu$ops \\ \hline
load/store queue & 32+32 entries \\ \hline
DL1 cache & 32KB, 2-way assoc. \\ \hline
IL1 cache & 16KB, 2-way assoc. \\ \hline
L2 cache & 256KB, 2-way assoc. \\ \hline
prefetcher & stride pref. (L1), stream pref. (L2) \\ \hline
page size & 4MB \\ \hline
SPM size & 216KB (up to 30 snapshots supported) \\ \hline
SPM throughput & 64 Bytes/cycle R/W \\ \hline
  \end{tabular}
      \caption{Baseline microarchitecture model. }
      \label{table:baseline}
\end{table}

The second benchmark is a real-world library \texttt{djpeg}, which is an application from the \emph{libjpeg} library that converts JPEG images into one of PPM, GIF, and BMP~\footnote{JPEG stands for Joint Photographic Experts Group, GIF stands for Graphics Interchange Format, PPM stands for Portable Pixmap Format, and BMP represents Device Independent Bitmap.}. The secret value for this benchmark is the input array that holds the image by representing the color and intensity of each pixel.  The core of the processing involves conditional branches that depend on each input array element. In contrast to the crypto library for which only tiny data (the key) is secret, input array in \texttt{djpeg} is substantially larger, e.g., a high-resolution photograph. 

The three output file types (PPM, GIF, and BMP) differ in  the number and type of instructions that are independent of the secret image.  Even the number of secret-dependent instructions is not the same due to the different number of decode steps each file type has.  The overall impact on the memory and the execution time depends on the output file type, so we use them as three separated workloads for the following analysis.

We compare \gls{sempe} against de facto technique for \gls{sdbcb} elimination: \gls{cte}. Specifically, we choose \gls{cte} version of the microbenchmarks written using \gls{fact}~\cite{cauligi_fact:_2017,Cauligi:2019:FDT:3314221.3314605}. \Gls{fact} offers a domain-specific language which greatly simplifies microbenchmark conversion to \gls{cte}. However, we did not apply \gls{fact} to \texttt{djpeg} because \gls{fact} has many limitations that prevents this, e.g. supporting only boolean and integers, lack of memory allocator support, lack of support for function pointers and lack of support for global variable (macros or multiple file inclusion). It took us approximately three weeks to convert the microbenchmark using \gls{fact}, which is a substantial programming effort. In contrast, with \gls{sempe}, the programmer only needs to insert directives into the code that specify the secret.

The benchmarks were compiled with clang/llvm on Debian GNU/Linux,
separating the secret-dependent code into its own compilation unit.  The secret-dependent code was compiled with optimizations disable to ensure that optimization does not inadvertantly reintroduce a side channel.
The rest of the benchmark code used the default optimization level, i.e., \verb'-O2'.

Each \gls{sblock} was manually instrumented with sJMP and eosJMP instructions.  Local variables were manually privatized (\gls{shadow}) as described in Section~\ref{sec:memory}, adding additional local variables and inline assembly to use \texttt{CMOV} after the secret branches. Both register allocated and memory allocated variables are privatized, so we can consider the worst case. The ArchRS mechanism described in Section~\ref{sec:register} allow to limit privatization and \texttt{CMOV}s to memory allocated variables only.

For baseline architecture, we model a processor with parameters shown in Table~\ref{table:baseline}. We use gem5 simulator~\cite{Binkert11thegem5} with an out-of-order processor configured  similar to the Intel Haswell~\cite{corporation_intel_2016} microarchitecture. The baseline differs from recent microarchitectures in terms of the cache size, to adjust for the benchmarks' smaller working set. 

We used a Scratchpad Memory that supports up to 30 register snapshots (one for each nested sJMP). Each snapshot contains two architectural register states and two bit-vectors with one bit per register each (Figure~\ref{fig:pipeline_ooo_with_spm}). Each register state contains the 48 architectural registers~\cite{noauthor_amd64_nodate}. The total size of a snapshot for each SecBlock is 7392 bytes. 


%% file: Text/evalu.tex
\section{Evaluation Results}
\label{sec:evaluation}


\subsection{Real-World Application Results}

\begin{figure}[t]
\centering
\includegraphics[width=1\linewidth]{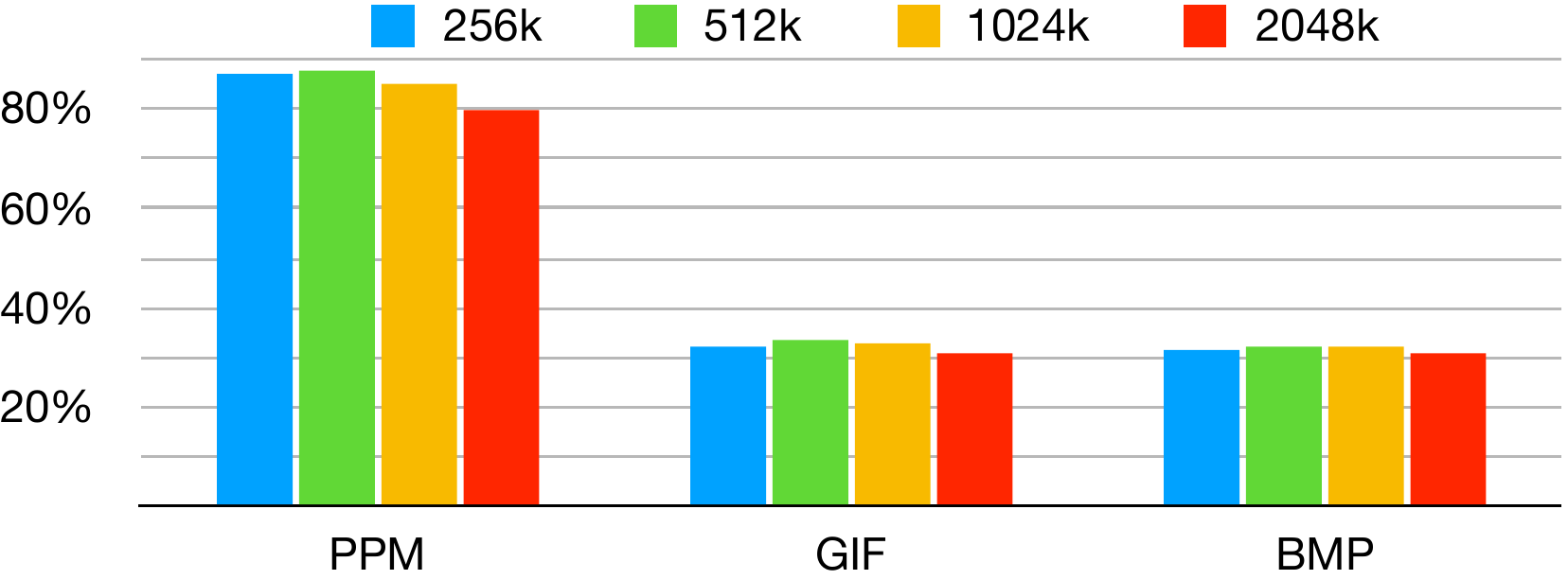}
  \caption{Execution time overhead for libjpeg with different image output format, varying input size.}
  \label{fig:libjpeg_execution}
\end{figure}

To evaluate \gls{sempe} performance, we display its execution time overheads over the baseline architecture that has no security protection (Figure~\ref{fig:libjpeg_execution}), for three output formats and input file sizes. The figure shows that the overheads vary between 31\% and 87\% across image output formats, but are not much affected by different image sizes. The overheads are much smaller than $2\times$, because the secure region only contributes to a fraction of the total instruction count. This factor also explains the variation across image output formats: the secure region in PPM contributes to much higher instruction count than GIF and BMP. On the other hand, the size of the input image does not affect the instruction count in the secure region because, on \texttt{djpeg}, the input array is decomposed into blocks, and each block performs several decompression steps depending on the type of output file-type produced. The \gls{sempe}  affects only the execution of \glspl{sblock}, allowing a consistent behavior that is largely independent of the input size.


The pipeline drain described in Figure~\ref{fig:pipeline_ooo_with_spm} produces \enquote{holes} in the pipeline, similar to a branch misprediction. This tends to increase \gls{cpi}. Factors that tend to decrease \gls{cpi} include not having branch misprediction (the branch predictor is not used for sJMP) and parallelism increases between branch paths due to the use of \gls{shadow}. Executing both branch paths may increase or decrease cache spatial and temporal locality. If the total working set of both branch paths increases beyond the cache capacity, locality may decrease and cache miss rates increase. On the other hand, if the working set of both paths overlap substantially, executing one path produces \emph{prefetching effect} for the alternate path, accelerating it. Apart from \gls{shadow} for privatization of local variables written in branch paths and used outside the secure region, different branch paths of a secure branch share all other memory locations, which improves the cache temporal locality. 

 \begin{figure*}[thbp]%
 \centering
 \begin{subfigure}{0.32\textwidth}
 \centering
 \includegraphics[width=1\linewidth]{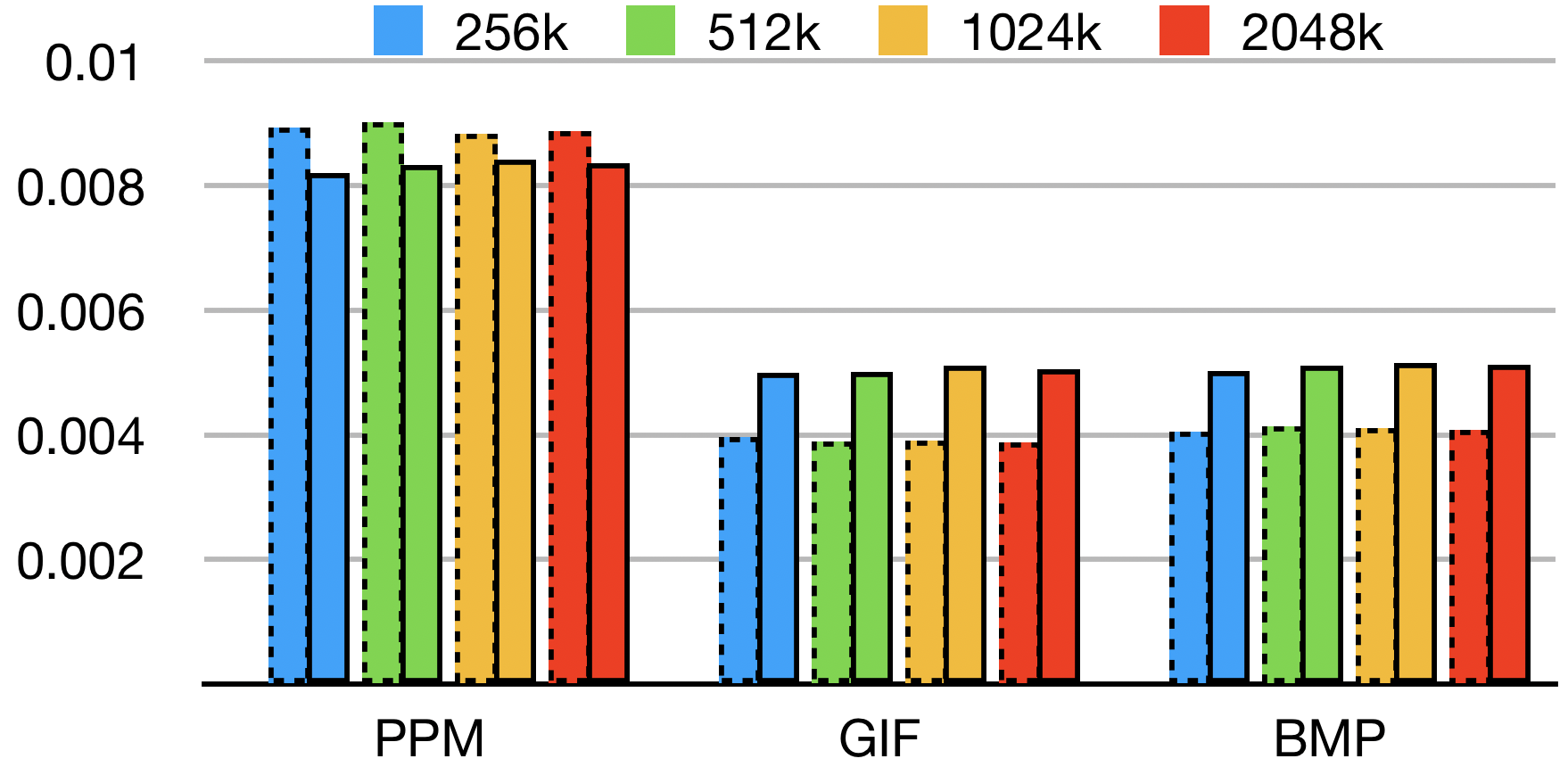}
  \caption{Instruction Cache}
   \label{fig:libjpeg_cache_icache}
 \end{subfigure}%
 \hfill
 \begin{subfigure}{0.32\textwidth}
 \centering
 \includegraphics[width=1\linewidth]{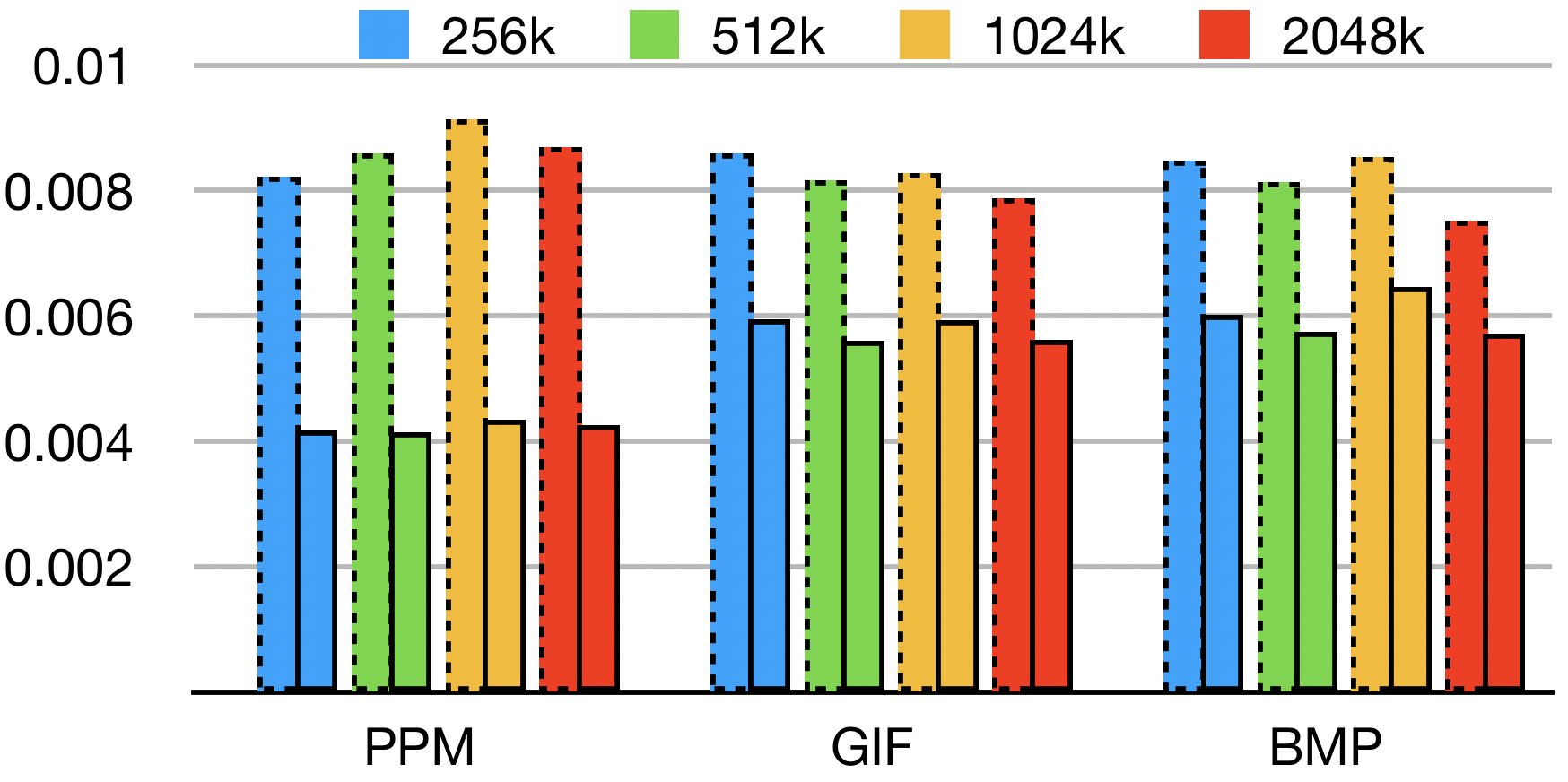}
  \caption{Data Cache}
   \label{fig:libjpeg_cache_dcache}
 \end{subfigure}%
 \hfill
 \begin{subfigure}{0.32\textwidth}
 \centering
 \includegraphics[width=1\linewidth]{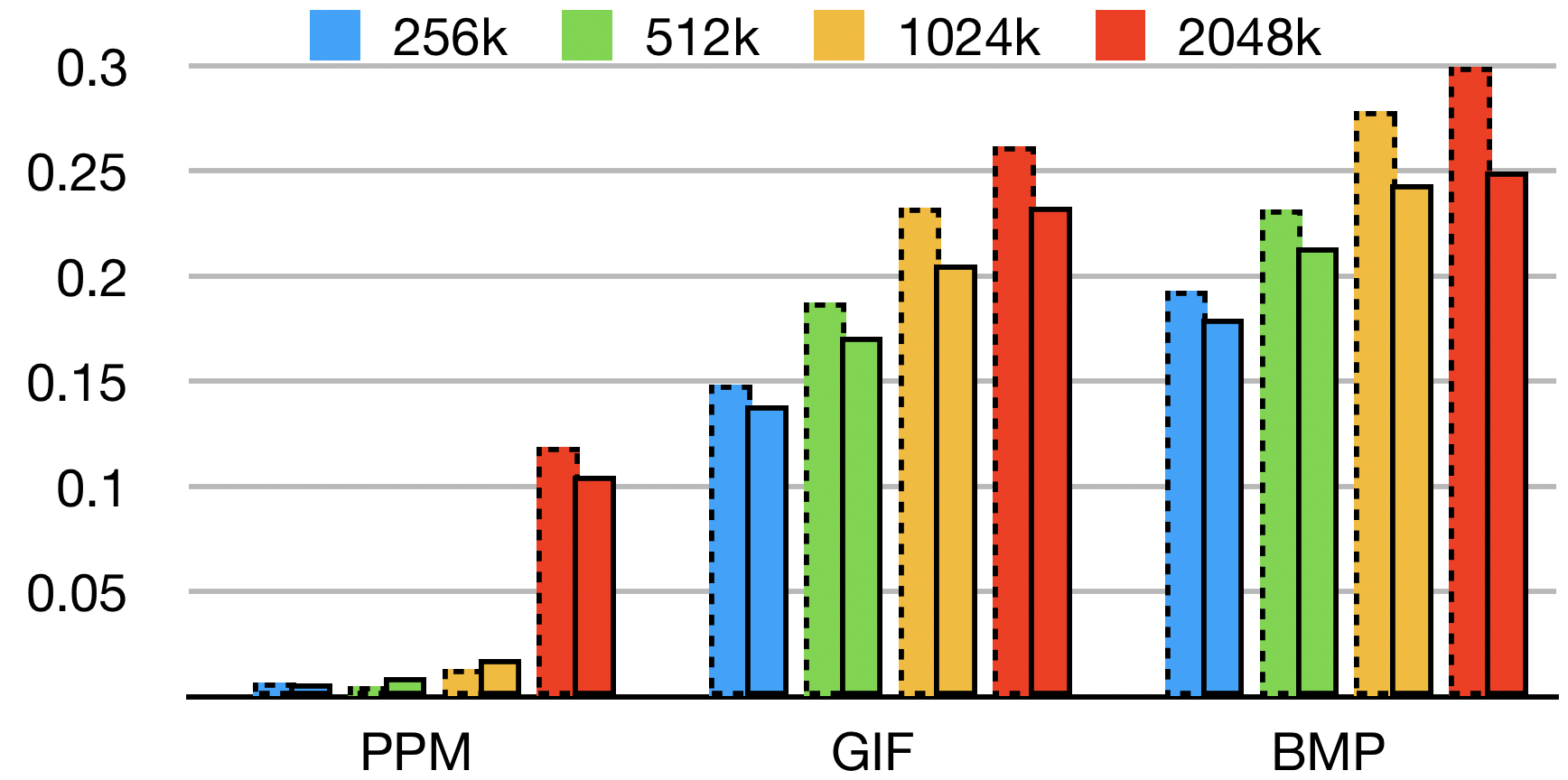}
  \caption{L2 Cache}
   \label{fig:libjpeg_cache_l2cache}
 \end{subfigure}%
\caption{Cache miss rates. Group of 2 columns: baseline (left, dashed line) and  \gls{sempe} (right, solid line). Lower is better. }
 \label{fig:libjpeg_cache}
 \end{figure*}
 
We analyze the impact of \gls{sempe} on cache memory in Figure~\ref{fig:libjpeg_cache}, which shows miss rates of the \gls{icache}, \gls{dcache}, and the \gls{l2}, across image output formats and image sizes for each format. Observe that the impact on instruction cache is unrelated to the size of the input image. \texttt{djpeg} divides the input into multiple sub-blocks, and the decompression work-flow is applied to each sub-block. The image size has an impact on the total number of \glspl{sblock} executed, but not on the number of instruction executed within a given \gls{sblock}. The \gls{icache} miss rate is low overall.  Despite the reduced \gls{icache} size used in our simulations, it is enough to contain the instructions that need to be fetched.

The situation changes when we dig into the \gls{dcache} miss rate analysis, shown in  Figure~\ref{fig:libjpeg_cache_dcache}. The two  \glspl{sblock} within a single decoding step of \texttt{djpeg} are, in all cases, small enough to fit the \gls{dcache}. The \gls{shadow} used during the \gls{sempe} play a fundamental role to take advantage of the principle of locality described earlier. Despite the execution of all the path of the sJMP, each path works on memory allocated (by the compiler) very closed each other. This memory is just a copy of the memory allocated before the secure region, that will be written only after the eosJMP by the \texttt{CMOV} instruction. The benefits of the \gls{dcache} miss-rate for \gls{shadow} have, as a consequence, a relevant  impact on the already low global miss rate. We perform a similar analysis on the \gls{l2} miss rates (Figure~\ref{fig:libjpeg_cache_l2cache}). The \gls{l2} miss rates are overall higher than for the data cache. However, the miss rates exhibit similar behavior as ones from the \gls{dcache}, even if this time changing the output file type, and consequently the number and type of instructions executed outside the \glspl{sblock}, has a much bigger impact on the total miss rates. 

\subsection{Microbenchmarks Results}
\label{microbench_analysis}

\begin{figure}[thbp]
\centering
  \includegraphics[width=1\linewidth]{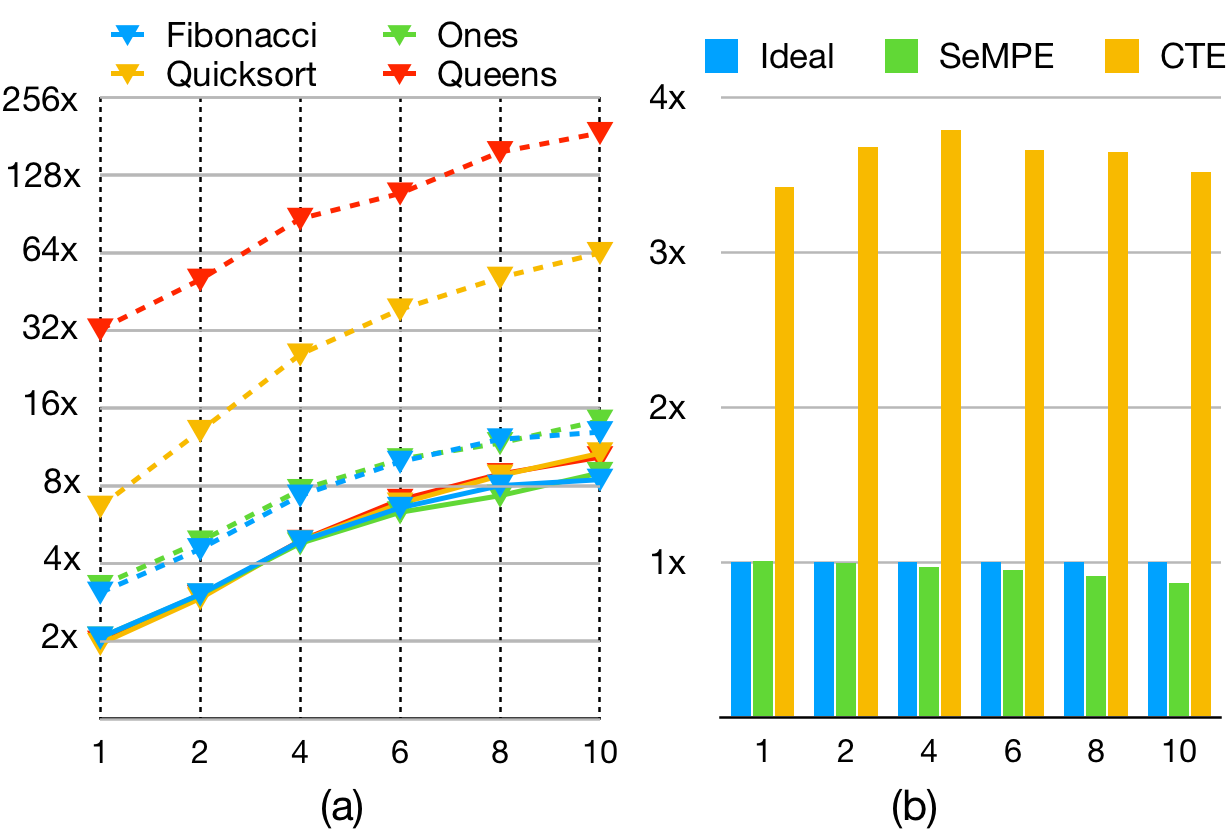}
  \caption{Execution time overheads affected by the nesting depth $W$ (X-axis): (a) SeMPE slowdown (solid line) vs. the slowdown due to CTE using FaCT (dashed line), and (b) Average slowdown normalized to ideal case.}
  \label{fig:trend}
\end{figure}

\gls{sempe} execute both paths of a secret-dependent branch, with instructions in each branch path unmodified, except for prefixing sJMP and inserting eosJMP instructions. Hence, we expect that the execution time with \gls{sempe} should be roughly linearly proportional to the number of branch paths executed. While other factors affect the execution time as well, such as the overhead of the multi-path jump back mechanism and the improvement obtained by the better usage of cache memories, they also scale proportionally to the number of branch paths executed. In contrast, we cannot expect that the execution time of \gls{cte} to be linearly proportional to the number of branch paths in the original program. \gls{cte} requires each statement in a branch path to be modified to include the unrolling of all the expressions that were part of the conditional statements (Figure~\ref{constanttime}). Hence, not only all statements in all branch paths are executed, but each statement takes longer to execute. Furthermore, the deeper the nesting level, and the more complex each conditional expression,  the more expressions need to be unrolled and the more complex each statement becomes. 
The execution time ratio to baseline for the microbenchmarks shown in  Figure~\ref{fig:trend}a confirm these expectations. The trends from no nesting ($W=1$) to deep nesting ($W=10$) are shown in Figure~\ref{fig:trend}a for \gls{sempe} (solid line)  and \gls{fact} (dashed line). Firstly, \gls{sempe} shows much lower overheads vs. \gls{cte} (note that the y-axes is in logarithmic scale). Furthermore, since each statement becomes more complex and translates to a higher instruction count, the execution time overheads are higher when the original code has more instructions. The slowdown of \gls{fact}, with $W=1$, ranges from $3\times$ for Fibonacci to $32\times$ for Queens (Figure~\ref{fig:trend}a).

Figure~\ref{fig:trend}a also shows that as the nesting depth is increased up to ten, execution time  slowdown increase for both \gls{sempe} and \gls{cte}. When $W=10$, \gls{sempe} increases execution time by roughly \mbox{$8.4-10.6 \times$}, consistent with the total number of branch paths of 11. \gls{cte}, on the other hand, slows down the execution between \mbox{$12.9-187.3\times$} (Figure~\ref{fig:trend}a). Such slowdowns render \gls{cte} impractical for use in user code which, unlike crypto library, may be executed frequently. Overall, \gls{cte} can be up to $18\times$ slower than \gls{sempe}. This is on top of \gls{cte}'s substantially higher programming effort. 
To remove side channel leakage from secret-dependent-conditional-branches, the execution must be indistinguishable for any secret value. Thus, unless two paths can be merged, the ideal overhead is the sum of execution time of all paths, which is exponential to nesting depth. \gls{sempe} beats this ideal overhead thanks to the \emph{prefetching effect}, hence it is low vs. ideal (Figure~\ref{fig:trend}b). In contrast, \gls{cte}’s overheads are super-exponential.

%% file: Text/concl.tex
\section{Conclusion}
\label{sec:conclution}

We introduced a hardware/software approach, SeMPE, that eliminates SDBCB without incurring high performance overheads or requiring high programming effort. SeMPE allows programmers to annotate secret branches in their program.  The \gls{isa} support is backward compatible. The architecture when encountering the new branch instruction executes both paths of the branch (one after the other) without consulting the branch predictor, thereby preventing the adversary from inferring secret from the executed path. SeMPE requires secret branches to be tracked using a hardware table that is small and simple (e.g. using \gls{lifo} instead of random access structure), and a small Scratchpad Memory to avoid the false register dependences that occur between both paths of a secure branch.

Hardware changes allow \gls{sempe} code to run on processor supporting multipath securely, or running on processor not supporting multipath fast. With hardware support, \texttt{CMOV} is only needed for phantom memory (but not register) dependences.
Code complexity of crypto code is low, so we evaluated \gls{sempe} using a real world application and microbenchmarks. 
We shown that the execution time with \gls{sempe} is near ideal; it increases linearly with the number of secret branch paths, independent from the size of workload executed in the \gls{sblock}. Our experiments also show a slight positive cache locality benefit from multi-path execution. When compared against \gls{cte} code derived using the state of the art \gls{cte} language and compiler (\gls{fact}), \gls{sempe} outperforms \gls{cte} substantially, by a factor of $1.6-18\times$.